\documentclass[prl,aps,showpacs,twocolumn]{revtex4}

\usepackage{epsfig}
\usepackage{epsf}
\usepackage{bm}                 
\usepackage{amsmath}
\usepackage{amssymb}
\usepackage{dcolumn}
\usepackage{graphicx}
\usepackage{psfrag}

\begin{document}


\title{Communicating continuous quantum variables between different Lorentz frames}

\author{Pieter Kok} \email{pieter.kok@hp.com}

\affiliation{Hewlett Packard Laboratories, Filton Road Stoke Gifford, 
Bristol BS34 8QZ, UK}

\affiliation{Quantum Computing Technologies Group, Jet Propulsion 
Laboratory, California Institute of Technology \\ Mail Stop 126-347, 
4800 Oak Grove Drive, Pasadena, California 91109-8099}

\author{Timothy C.\ Ralph} 

\affiliation{Department of Physics, Centre for Quantum Computer Technology, 
University of Queensland, St.\ Lucia, Queensland 4072, Australia}

\author{Gerard J.\ Milburn} 

\affiliation{Department of Physics, Centre for Quantum Computer Technology, 
University of Queensland, St.\ Lucia, Queensland 4072, Australia}

\date{\today}

\begin{abstract}
 We show how to communicate Heisenberg-limited continuous (quantum)
 variables between Alice and Bob in the case where they occupy two
 inertial reference frames that differ by an unknown Lorentz
 boost. There are two effects that need to be overcome: the Doppler
 shift and the absence of synchronized clocks. Furthermore, we show
 how Alice and Bob can share Doppler-invariant entanglement, and we
 demonstrate that the protocol is robust under photon loss.
\end{abstract}


\maketitle

\noindent
Quantum communication spans a wide range of topics, from cryptography
and teleportation, to multi-party entanglement protocols, error
correction and purification \cite{bouwmeester00}. Often, the emphasis
is on how we can outperform classical communication protocols with
shared entanglement, using only local operations and classical
communication. However, we have to be very careful when we consider
sharing entanglement, since its distribution is a physical process. As
a consequence, it is subject to uncertainties and noise. Moreover,
quantum communication protocols typically assume that all parties have
perfect knowledge about a global frame of reference. In other words:
they all agree on which way they call ``up''. Both the distribution
and the local (re-) definition of the quantum states needs to be
addressed in any practical implementation of the communication
protocol.

When Alice and Bob want to establish a (quantum) communication
channel,  they first need to agree on the specific protocols that they
are going to use. As was suggested above, one also might think that
they need to share special information such as a (global) fixed frame
of reference (see, for example, Ref.~\cite{rudolph03} and references
therein), or synchronized clocks. However, there are quantum
communication protocols  that can circumvent the need for, e.g., a
global frame of reference \cite{bartlett03}. On the other hand,
establishing perfect clock synchronization has proved much trickier
\cite{jozsa00,preskill00}. The question we address here is whether
Alice and Bob can communicate continuous (quantum) variables when they
have no prior information about their respective inertial frames of
reference. Also, the case for discrete variables was recently proposed
\cite{bartlett04}. 

Such a problem obviously needs to be formulated in a relativistic
setting.  Recently, there has been considerable interest in
relativistic quantum information. It was shown that a fundamental
information-theoretic concept such as entropy is not a relativistic
scalar \cite{peres02} and a Lorentz transformation of subsystems mix
the entanglement between spin and momentum
\cite{gingrich02}. Furthermore, the relativistic transformation of
Bell states was derived \cite{alsing02}. In this Letter, we look for
invariant quantum states with respect to Lorentz boosts. In addition,
we will construct entanglement between the frames of Alice and Bob.

Suppose that Alice and Bob occupy different inertial frames of
reference. If they wish to communicate a real number $\lambda$ it is
natural to use electromagnetic (quantum) waves, because of its robust
properties for long-distance communication. If Alice and Bob do not
know their relative velocities, the communication is hampered by two
effects: First, any signal sent from Alice to Bob (and vice versa)
will suffer from an unknown Doppler shift. Secondly, the local clocks
of Alice and Bob will run at different rates to an outside
observer. Since many quantum-optical measurements (such as, e.g.,
homodyne detection) rely intrinsically on timing information, we need
to remove the time dependence of the states in the relevant part of
the wave function. 

The classical way to communicate a real number $\lambda$ is for Alice to
send a pulse of coherent light and for Bob to measure the
intensity. The Doppler shift only changes the frequency, not the number
of photons, so the intensity is invariant. Similarly the time
dilation will stretch or compress the pulse, and provided Bob
integrates over the whole pulse, he will get the same result. If
the coherent amplitude of the pulse is made sufficiently large then
the photo-detector will self-homodyne and hence produce a continuous
spectrum (corresponding to the in-phase quadrature) on which to encode
$\lambda$.

However, coherent light is ultimately shot-noise limited, i.e., we can
only estimate $\lambda$ up to a precision $\sqrt{\lambda}$. Secondly,
we are interested in the invariant subspaces of continuous
quantum-variable systems under Lorentz boosts. The protocol we propose
generates these invariant subspaces, and yields Heisenberg-limited
precision in determining $\lambda$. Furthermore, we will show that our
protocol is robust under photon loss.

Let the four-vector potential of the electromagnetic field be given by 
\begin{equation}
  A^{\mu}(x) = \int d^4 k \, \delta(k^2 - \vec{k}^2 + k_0^2) \sum_j \left[
  \epsilon_j^{\mu} \hat{a}_j (k) e^{ikx} + \mathrm{H.c.} \right] \; ,
\end{equation}
where $k$ and $x$ are the wave and position four vectors, and
$\epsilon_j$ is the $j$-polarization vector. The frequency of a field
mode is given by $k_0 \equiv \omega_k$. Both Alice and Bob describe the
field in the Coulomb gauge [$\nabla\cdot\mathbf{A}=0$, with $A^{\mu} =
(\phi,\mathbf{A})$], which is not Lorentz invariant. It was shown by
Kok and Braunstein \cite{kok04} that a pure Lorentz boost $\Lambda$
without rotation does not affect the polarization of the field in the
Coulomb gauge, and that the annihilation operator transforms as
$\hat{a}_j' (k) = \sum_l U_{jl}\; \hat{a}_l (\Lambda k)$ under Lorentz
transformations. The $SU(2)$ matrix $U$ corresponds to the overall
spatial rotation. Such rotations were considered by Bartlett {\em et
  al}. \cite{bartlett03}, hence we confine our discussion to pure
boosts, yielding $\hat{a}_j' (k) = \hat{a}_j (\Lambda k)$. This allows
us to suppress the polarization, and treat the electromagnetic field
as a set of scalar fields.   

The annihilation operator can be written as
\begin{equation}
  \hat{a}(k) = \sqrt{\frac{\omega_k}{2\hbar}}\, \hat{q}(k) +
  \frac{i}{\sqrt{2\hbar \omega_k}}\, \hat{p}(k)\; .
\end{equation}
Here, $\hat{q}(k)$ and $\hat{p}(k)$ are the (Hermitian) quadratures of 
the field mode $k$, and they obey the canonical commutation relation
$[\hat{q}(k), \hat{p}(k')] = i\hbar\, \delta(k-k')$. A Doppler shift
due to a Lorentz boost $\Lambda$ between inertial frames will result
in a transformation $\omega_k \rightarrow \mu^2 \omega_k$, where $\mu^2 =
\sqrt{1-\beta^2}$. Here, $\beta=v/c$ is Bob's velocity. 

The phase number $kx$ is a scalar (it is the number of wave crests
counted by an observer), and therefore invariant under
boosts. The modes are thus transformed according to 
\begin{equation}
  \hat{a}'(k) \rightarrow \sqrt{\frac{\omega_k'}{2\hbar}}\,
  \hat{q}(\Lambda k) +
  \frac{i}{\sqrt{2\hbar \omega_k'}}\, \hat{p}(\Lambda k)\; ,
\end{equation}
which is equivalent to the transformation
\begin{equation}\label{mu}
  \hat{q}(k) \rightarrow \hat{q}'(k) = \mu\, \hat{q}(\Lambda k)
  \quad\text{and}\quad
  \hat{p}(k) \rightarrow \hat{p}'(k) = \frac{1}{\mu}\, \hat{p}(\Lambda k)\; .
\end{equation}
A Doppler shift therefore clearly corresponds to a squeezing operation
in the phase space $(q,p)$. Indeed, the symmetry group of a quantum
oscillator with variable frequency is $SU(1,1)$ \cite{perelomov86}.

Measuring certain observables of the electromagnetic field often
implicitly assumes the existence of a local clock. For example,
homodyne detection uses a local oscillator, which serves as the
clock. When Alice prepares a state using a local oscillator, there is
a priori no reason to believe that Bob's measurement using a different
clock running at a different rate will project onto the same
state. The second physical hurdle we have to overcome in communicating
continuous variables between different inertial reference frames is
therefore to remove the time dependence of the state preparation and
measurement stages.

Locally, we can write the time evolution in terms of the following
$SU(2)$ quadrature transformation of the field modes:
\begin{eqnarray}\label{quad}
  \begin{pmatrix}
   \hat{q} \\
   \hat{p}
  \end{pmatrix}
  \rightarrow
  \begin{pmatrix}
   \hat{q}' \\
   \hat{p}'
  \end{pmatrix}
  =
  \begin{pmatrix}
    \cos\phi & \sin\phi \\
   -\sin\phi & \cos\phi
  \end{pmatrix}
  \begin{pmatrix}
   \hat{q} \\
   \hat{p}
  \end{pmatrix}\; .
\end{eqnarray}
These transformations must leave invariant the part of the quantum
state that encodes $\lambda$. However, this does not necessarily mean
that the entire quantum state is invariant \cite{knill00}. Indeed, we
will construct invariant states of infinite energy, whose regularized
finite-energy states are no longer completely invariant, but retain
sufficient invariance to faithfully encode $\lambda$.

The physical motivation for this discussion was that Alice wants to
communicate a real number $\lambda$ to Bob, using the quantum
properties of light. We can consider two distinct cases of quantum
communication of continuous variables. Firstly, we can use quantum
states to communicate a classical variable. Secondly, we can
communicate a continuous quantum variable, which includes
superpositions of real numbers. In the first part of this Letter we
consider the first case, and in the second part, we address the latter.

To communicate a real number $\lambda$ with Bob, Alice needs to
send a beam or pulse of light in a state $|\lambda\rangle$, such that
Bob can retrieve $\lambda$ with some (finite) precision
$\Delta\lambda$. In other words, the beam or pulse carrying the
information about $\lambda$ must be invariant under Doppler shifts and
local time translations. The simplest operator that obeys these
requirements is reminiscent of the angular momentum operator
$\hat{L}(k) = \hat{q}_1(k) \hat{p}_2(k) - \hat{p}_1(k) \hat{q}_2(k)$,
where the subscripts on $q$ and $p$ denote two distinct modes (if one
uses the polarization degree of freedom to distinguish these modes,
then $\hat{L}$ is also invariant under spatial rotations
\cite{bartlett03}). It exhibits the famous singlet structure of
$SU(2)$ representations. Singlets are invariant under unitary
transformations of the form $U \otimes U$ (where $U$ is an arbitrary
unitary transformation on the subsystem). More importantly,
$\hat{L}(k)$ is also invariant under $SU(1,1)$ group transformations,
and should therefore be the building block for constructing invariant
subspaces (so-called {\em decoherence-free subspaces}
\cite{zanardi97,lidar98})  for continuous-variables quantum
communication. We therefore need to construct the eigenstates
$|\lambda\rangle$ of $\hat{L}$:
\begin{equation}
 \hat{L}(k) |\lambda\rangle = \hat{L}(\Lambda k) |\lambda\rangle =
 \left( \hat{q}_1 \hat{p}_2 - \hat{p}_1 \hat{q}_2 \right)
 |\lambda\rangle =  \lambda |\lambda\rangle\; . 
\end{equation}
From now on, we implicitly assume that $q_j$ and $p_k$
are labeled by $k$ and $\Lambda k$ for Alice and Bob respectively.

\begin{figure}[t]
  \begin{center}
  \begin{psfrags}
     \psfrag{q1}{$q_1$}
     \psfrag{q2}{$q_2$}
     \psfrag{Re}{Re}
     \psfrag{Im}{Im}
     \psfrag{j}{$\lambda=3$}
     \psfrag{i}{$\lambda=5$}
        \epsfig{file=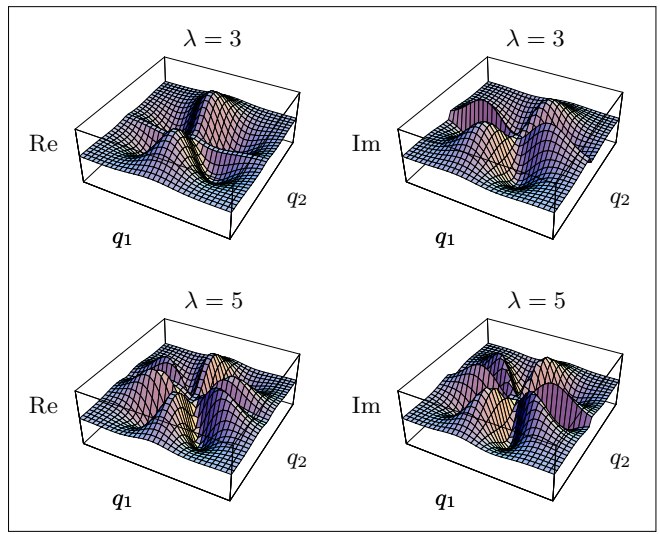, height=7cm}
  \end{psfrags}
  \end{center}
  \caption{The real and imaginary parts of the probability amplitude for
  the quadrature phases with finite energy $|\psi_{\lambda}^a\rangle$
  with $a=1$, $\lambda=3$ and $\lambda=5$ on the interval $-3 \leq
  q_1, q_2 \leq 3$.}   
  \label{fig1}
\end{figure}

Next, we seek the eigenstates that facilitate the communication of a 
continuous variable $\lambda$ between Alice and Bob. Define 
$\psi_{\lambda}(q_1,q_2) \equiv \langle q_1, q_2|\lambda\rangle$ as
the probability amplitude for the quadrature phases $q_1$ and $q_2$, 
and $\hat{p}_j = i\hbar\, \partial/\partial q_j$. Then the eigen 
equation is given by 
\begin{equation}
  i\hbar \left( q_1 \frac{\partial}{\partial q_2} - q_2 
  \frac{\partial}{\partial q_1} + \frac{i\lambda}{\hbar} \right) 
  \psi_{\lambda}(q_1,q_2) = 0\; ,
\end{equation}
which gives us the (unnormalized) state
\begin{equation}\label{state}
  \psi_{\lambda}(q_1,q_2) \propto \exp\left[\frac{i \lambda}{\hbar}
  \arctan\left(\frac{q_1}{q_2}\right) \right]\, (q_1^2 + q_2^2)\; .
\end{equation}
The task is now for Alice to send a state of the form of
Eq.~(\ref{state}) to Bob, who can then retrieve the value $\lambda$ by
measuring the operator $\hat{L}(\Lambda k) = \hat{q}_1 \hat{p}_2 - \hat{p}_1
\hat{q}_2$. However, the state $\psi_{\lambda}(q_1,q_2)$ has
infinite mean energy, and as a consequence, Alice cannot send this
exact state. We can regularize the `ideal' state of Eq.~(\ref{state}) 
such that they are no longer states with infinite energy (see
Fig.~\ref{fig1}): 
\begin{equation}\label{state2}
  \psi_{\lambda}^a (q_1,q_2) \propto 
  (q_1^2 + q_2^2) e^{\frac{i\lambda}{\hbar} 
  \arctan \left( \frac{q_1}{q_2} \right) -a (q_1^2 + q_2^2)/2}\; .
\end{equation}
The exponential factor $\exp[-a(q_1^2 + q_2^2)/2]$ (with $\mathrm{Re}\,
a > 0$) ensures that the wave function $\psi_{\lambda}^{a} (q_1,q_2)$
remains finite and localized in phase space. Indeed, with the
normalization constant $\sqrt{a^3/2\pi}$,  we find that the mean
energy $\langle E \rangle$ of a frequency  mode $\omega_k$ is
\begin{equation}
  \langle E \rangle = \hbar\omega_k \left( \frac{1}{2}+\frac{3}{a}
  \right) \; ,
\end{equation}
where we used that $E = \hbar\omega_k (\hat{n}_1 + \hat{n}_2)$, and
$\hat{n}_i = q_i^2 - \partial^2_{q_i}$. It is immediately  clear that
the mean energy diverges only if $a$ becomes zero and
$\psi_{\lambda}^{a}$ becomes $\psi_{\lambda}$. For finite $a$ the
dispersion in the energy $\Delta E$ is also finite.

The state $\psi_{\lambda}^a(q_1,q_2)$ is not strictly invariant under
Lorentz boosts. In particular, a boost parametrized by $\mu$ in
Eq.~(\ref{mu}) leads to the transformation $a \rightarrow \mu^2
a$. This corresponds to an expected shape change in the wave
packet. However, it is easily shown that Bob's measurement of the
observable $\hat{L}(\Lambda k)$ is not affected by this
transformation. When we calculate the expectation values of
$\hat{L}(\Lambda k) = q_1 \partial_{q_2}  - q_2 \partial_{q_1}$ and
$\hat{L}^2$, we find 
that $\langle \hat{L}(\Lambda k) \rangle = \lambda$ and $\langle
\hat{L}^2(\Lambda k)\rangle=\lambda^2+1$. The error  in $\lambda$ is then
$(\Delta \lambda)^2 = \langle \hat{L}^2\rangle -  \langle
\hat{L}\rangle^2  = 1$ in units of $\hbar$. That is, Bob can in
principle retrieve the value of $\lambda$ by measuring the observable
$\hat{L}$ without Alice having to resort to infinite energy states. In
practice, Bob will induce an error $\delta\lambda$ associated with his
measurement scheme.

Next, we consider what happens when a photon is lost in the process
of sending the state $\psi_{\lambda}^a$. The loss of a photon is
modeled by a beam-splitter $\hat{B}$ with reflection amplitude
$\kappa\in\mathbb{R}$, where 
the reflected beam is traced over. Consequently, $\kappa^2$ is the
photon loss. When the loss is small, we only need to take into account
the first few terms of the unitary evolution $U =
\exp[i\kappa\hat{B}]$:
\begin{equation}\nonumber
 \psi_{\lambda,\rm out}^a (q_1,q_2,q_3) = \left( 1 + i\kappa\hat{B}
 -\frac{\kappa^2}{2} \hat{B}^2 \right) \psi_{\lambda}^a (q_1,q_2) \; .
\end{equation}
In the Bargmann representation, we have $\hat{B} = i q_3
\partial_{q_1} - i q_1 \partial_{q_3}$. When we collect terms up to
$\kappa^2$, the expectation value of an observable $\hat{A}$ in the
presence of the loss mechanism is given by
\begin{eqnarray}
 \langle\hat{A}\rangle_B &=& \langle\hat{A}\rangle + \kappa^2 \int
 dv\, \bar{\psi}_{\lambda}^a \hat{B}^{\dagger}\hat{A}\hat{B}\,
 \psi_{\lambda}^a \cr 
 && + \frac{\kappa^2}{2} \int dv\, \bar{\psi}_{\lambda}^a (
 \hat{B}^{\dagger 2} \hat{A} + \hat{A}\hat{B}^2 )\, \psi_{\lambda}^a \; ,  
\end{eqnarray}
where the integral $\int dv = \iiint dq_1\, dq_2\, dq_3$, and
$\bar\psi$ denotes the complex conjugate.

When we calculate $\langle\hat{L}\rangle_B$ and
$\langle\hat{L}^2\rangle_B$, we find
\begin{equation}
 \langle\hat{L}\rangle_B = \lambda \qquad\text{and}\qquad 
 \Delta L_B \simeq 1 + \frac{a\kappa^2}{12} (2+\lambda^2)\; .
\end{equation}
The expectation value $\langle\hat{L}\rangle_B =
\langle\hat{L}\rangle$ is not affected, while the precision
$\Delta L_B$ starts to deteriorate when $\lambda$ becomes too
large. However, this can be made arbitrarily small by reducing
$a$. The ideal infinite-energy states ($a\rightarrow 0$) do not suffer
from reduced precision at all.

The states in Eqs.~(\ref{state}) and (\ref{state2}) are in some sense
``ideal'' choices, but at this point we have no idea how to produce
these states. Furthermore, they are by no means the only choice. Any
polynomial operator $\hat{A} = \hat{A}(\hat{L})$ that  has sufficient
structure to encode $\lambda$ must be invariant under the  quadrature
transformations of Eq.~(\ref{quad}), and the eigenstates of  $\hat{A}$
are therefore suitable for our purposes: $\hat{A}(\hat{L})\,
|\Phi_{\alpha}\rangle = \alpha(\lambda)\, |\Phi_{\alpha}\rangle$,
where $\alpha(\lambda)$ is a bijective function of $\lambda$. The set
of all operators $\hat{A}$ with their associated detection of
$\alpha(\lambda)$ determine a class of possible protocols to
communicate between different inertial reference frames. An important
operator of this type is given by $\hat{A} = \exp[i\lambda (\hat{q}_1
\hat{p}_2 - \hat{q}_2 \hat{p}_1)]$. This operator can in principle be
generated with parametric down-conversion, and it opens a gateway to
the practical implementation of this protocol. To this end, we need to
find optimal ways to estimate $\lambda$ \cite{bollini93,milburn94}.

In the remainder of this Letter, we consider superpositions of the
form $\int d\lambda f(\lambda) |\lambda\rangle$, where $\int d\lambda
|f(\lambda)|^2 =1$. Since Lorentz transformations are unitary, these
superpositions remain coherent when sent to Bob. However, the
superposition might change due to the Lorentz transformation. 

Consider $\hat{L}$ as the generator of a symmetry group parametrized
by a conjugate variable $\beta$: $|\psi(\beta+d\beta)\rangle =
\exp(id\beta\hat{L}/\hbar) |\psi(\beta)\rangle$. Using a Taylor
expansion in $\beta$, we find
\begin{equation}
 i\hbar \frac{d}{d\beta} |\psi(\beta)\rangle = \hat{L}
 |\psi(\beta)\rangle\; ,
\end{equation}
or $|\psi(\beta)\rangle = \exp(i\beta\hat{L}/\hbar)
|\psi\rangle$. In the interaction picture, we write
$|\psi\rangle$ independent of $\beta$, and an operator $\hat{A}$
evolves according to $\exp(i\beta\hat{L}/\hbar) \hat{A}
\exp(-i\beta\hat{L}/\hbar)$. 

General superpositions of $|\lambda\rangle$ then evolve according to 
\begin{equation}
 \int d\lambda f(\lambda) |\lambda\rangle ~\rightarrow~ \int d\lambda
 f(\lambda) e^{-i\lambda\beta/\hbar} |\lambda\rangle\; .
\end{equation}
Given a suitable superposition, Alice and Bob can determine $\beta$
with (possibly Heisenberg limited) precision. Alternatively, Alice can
send four-mode superpositions of the form
\begin{equation}
 \int d\lambda f(\lambda) |\Phi_{\lambda}\rangle \equiv \int d\lambda
 f(\lambda) |\lambda,-\lambda\rangle\; , 
\end{equation}
which are manifestly invariant under $\beta\rightarrow\beta'$. We can
now build entangled states that are invariant under Lorentz
transformations:
\begin{equation}
 |\Upsilon\rangle_{AB} = \int d\lambda f(\lambda)\,
 |\Phi_{\lambda}\rangle_A |\Phi_{\lambda}\rangle_B\; .
\end{equation}
A complete set of continuous-variable ``Bell''-states is then given by
$\int d\lambda\, g(\lambda,\zeta)\, |\Phi_{\lambda}\rangle_A
|\Phi_{\lambda\oplus\zeta}\rangle_B$ \cite{braunstein98}, where
$\zeta\in\mathbb{R}$, and the subscripts $A$ and $B$ denote the modes
held by Alice and Bob respectively. At this point, we should note that we
can no longer use the polarization degree of freedom alone to distinguish
between $q_j$ and $q_k$. However, we {\em can} use a specific color
ordering, since Doppler shifts do not change the order of the spectrum.

In conclusion, we have constructed a class of finite energy states of
the electromagnetic field that can be used to send continuous quantum
variables from one Lorentz frame to another without knowledge of their
relative velocity. These states are therefore invariant under Doppler
shifts. Also, the wave function needs to have a time-independent
component since time dilatation prevents Alice and Bob from having
synchronized local oscillators. We found a multi-mode operator that
may serve as a natural building block to construct invariant subspaces
under Lorentz transformations over continuous variables. This protocol
can also be used by Alice and Bob to share continuous-variable
entanglement. Furthermore, the states are robust under small photon
loss.

P.K.\ wishes to thank Sam Braunstein and Bill Munro for valuable
comments. Part of the research described in this Letter was carried out
at the Jet Propulsion Laboratory, California Institute of Technology,
under a contract with the National Aeronautics and Space
Administration (NASA). P.K.\ was supported by the United States
National Research Council, the Australian Centre for Quantum Computer
Technology, and the European Union RAMBOQ project.

\bibliography{frame}

\end{document}